\documentstyle[11pt]{article}
\begin{document}



\def\e{{\hat e}}
\def\ie{{$i.e.$}}
\def\ra{{\rightarrow}}
\def\a{{\alpha}}
\def\b{{\beta}}
\def\eps{{\epsilon}}
\def\n{{\eta}}
\def\g{\gamma}
\def\s{{\sigma}}
\def\r{{\rho}}
\def\z{{\zeta}}
\def\x{{\xi}}
\def\d{{\delta}}
\def\t{{\theta}}
\def\T{{\Theta_A}}
\def\l{{\lambda}}
\def\ca{{\cal A}}
\def\cc{{\cal C}}
\def\cd{{\cal D}}
\def\ce{{\cal E}}
\def\cg{{\cal G}}
\def\ci{{\cal I}}
\def\co{{\cal O}}
\def\cn{{\cal N}}
\def\car{{\cal R}}
\def\cs{{\Theta_S}}
\def\cu{{\cal U}}
\def\cz{{\cal Z}}
\def\cv{{\nu}}
\def\ck{{\cal K}}
\def\pr{{\partial}}
\def\prt{{\pr}_{\t}}
\def\tri{{\triangle}}
\def\na{{\nabla }}
\def\S{{\Sigma}}
\def\G{{\Gamma }}
\def\sp{\vspace{.1in}}
\def\hs{\hspace{.25in}}

\newcommand{\be}{\begin{equation}} \newcommand{\ee}{\end{equation}}
\newcommand{\bea}{\begin{eqnarray}}\newcommand{\eea}
{\end{eqnarray}}


\begin{titlepage}

\begin{flushright}
{{G\"oteborg-ITP-99-17}\\{hep-th/9911251}\\
{November, 1999}\\
{\tt Nucl.Phys.B.} (in press)}
\end{flushright}

\begin{center}
\baselineskip= 24 truept
\vspace{.5in}

{\Large \bf Non-commutativity, Zero modes and D-brane Geometry}

\vspace{.4in}
{\large Supriya Kar\footnote{supriya@fy.chalmers.se}}

\end{center}
\begin{center}
\baselineskip= 18 truept

{\large \it Institute for Theoretical Physics \\
G\"oteborg University and Chalmers University of Technology \\
S-412 96 G\"oteborg, Sweden}

\vspace{.2in}

({\today})

\end{center}

\vspace{.4in}

\baselineskip= 15 truept

\begin{center}
{\bf Abstract}
\end{center}
\vspace{.25in}

The non-commutative geometry is revisited from the perspective
of a generalized $D$ p-brane. In particular, we analyze the
open bosonic string world-sheet description and show that an
effective non-commutative description on a $D$ p-brane
corresponds to a re-normalized world-volume. The world-volume
correlators are analyzed to make a note on the non-commutative
geometry. It is argued that the $U(1)$ gauge symmetry can be 
viewed in a homomorphic image space for the 
non-commutative space of coordinates with an $SO(p)$ symmetry.
In the large B-field limit, the non-commutativity reduces 
to that among the zero modes and the world-volume description 
is precisely in agreement with the holographic principle.

\vspace{.25in}

\thispagestyle{empty}
\end{titlepage}

\baselineskip= 18 truept

\section{Introduction}

Dirichlet ($D$) p-branes are non-perturbative \cite{polchinski} topological 
defects seen in string theory. They have been studied extensively to
explore the non-perturbative domain \cite{dkps,bfss,ikkt}
using mathematical tools familiar in 
string theory. Among various aspects of a generalized $D$ p-brane, the 
non-commutative geometry \cite{connes,hull} on its world-volume 
has significantly played a role to enhance our understanding of the
gauge theory. In a
recent analysis \cite{sw}, a precise relation between the non-commutative
and commutative gauge symmetries on a $D$ p-brane world-volume has been
established. 

\sp
\hs
The idea of non-commutativity is quite natural in the theory
of open strings. For instance, its world-sheet boundary is an one-dimensional
ordered space which is evident from its vertex operators insertion. In fact,
under a rotation of the world-sheet, the vertex operators retain the 
cyclic symmetry and implies a non-commutative geometry there.
In the context of open string field theory, 
the non-commutative geometry has also been addressed \cite{witten86}. Since
the dynamical aspect of a $D$ p-brane has its origin in open string physics,
interestingly, the longitudinal collective coordinates for a generalized 
$D$ p-brane also become non-commutative. This implies that the world-volume
geometry can be
analyzed perturbatively as a deformation to the commutative one. 
The (convergent) series expansion has been shown to be identical with the 
Moyal product following a boundary conformal field theory technique 
\cite{schomerus}. In the recent past, non-commutative
geometry in the realm of a $D$ p-brane have been explored in great details
by various authors \cite{berkooz}-\cite{okuyama}.
Further investigation in this direction is  
believed to provide a conceptual understanding of the D-brane 
geometry and may enlighten a formulation for the quantum gravity.

\sp
\hs
In this article, we focus on the non-commutative description on a
world-volume of a generalized $D$ p-brane with an 
emphasis on the zero modes. We consider the re-normalized world-volume
to bring a note on the effective non-commutativity as seen in a
physical process, such as scattering phenomena. We study the
world-volume correlators for the renormalized $D$ p-brane to comment
on the non-commutative and commutative symmetries. By diagonalizing 
the world-volume into two dimensional block diagonal forms, we
discuss the possible geometry on a rotated $D$ p-brane. Finally, the
large B limit is analyzed and the holographic correspondence is argued
to relate the non-commutative and commutative geometries. In the limit,
the world-volume becomes non-local due to the zero modes and  
ground state wave function reprsenting a
$D$-particle degenerates for the Landau level.

\sp
\hs
We plan to present the article as follows.
In section 2.1, we briefly discuss the $D$ p-brane dynamics
in the open string channel to obtain the boundary conditions. In sec. 2.2,
we present the non-commutative aspect of a $D$ p-brane. The
re-normalization of the world-volume is performed in sec. 2.3. Subsequently
in sec. 3.1, we obtain the world-volume correlators for a generalized
$D$ p-brane. Sec. 3.2, deals with the world-volume rotational symmetry
$SO(p)$ for a $D$ p-brane and the role of zero modes are discussed. The
large B-field limit is discussed in sec. 3.3. and we conclude
the article with discussions in sec. 4.

\section{Open string analysis}

\subsection{Generalized D-brane description}

Let us begin with the open bosonic string dynamics \cite{tseytlin,callan}
in presence of closed string background fields: metric ${\hat G}_{\mu\nu}$,
Kalb-Ramond (KR) field ${\hat B}_{\mu\nu}$
($\mu=0,1,\dots 25$) and a constant dilaton. 
In a world-sheet conformal
gauge, the open string action can be given by
\be
S\ = {1\over{4\pi\a'}} \ \int_{\S}
\left [\ {\hat G}_{\mu\nu} {\pr_{\bar a}} X^{\mu}
{\pr^{\bar a}} X^{\nu} \ -\ \eps^{\bar a\bar b}\ {\hat B}_{\mu\nu} 
{\pr_{\bar a}} X^{\mu} {\pr_{\bar b}} X^{\nu}\ \right ]
\ + \ \ {\int}_{\pr {\S}}
{A}_{\mu}\pr_{\tau} X^{\mu} \ ,\label{openaction}
\ee
where the second integral denotes the $U(1)$ gauge  
(field: ${A}_{\mu}$) interaction at the boundary, $\pr\S$.

\sp
\hs
Now consider an arbitrary $D$ p-brane{\footnote{For definiteness, consider 
an even $p$ case in type IIA
string theory. The odd $p$ case can as well be discussed for the
Euclidean world-volume geometry where it is even dimensional.}}
with $p$-spatial coordinates describing its world-volume in
the open string theory (\ref{openaction}). The background fields can be
seen to induce metric $h$ and the antisymmetric B-field on the
world-volume of the $D$ p-brane. In a gauge,
$h_{0i}=-1$ and $B_{0i}=0$ for ($i=1,2,\dots p$), they can be
given by
\bea
h_{ij}&=&{\hat G}_{\mu\nu}\ \pr_iX^{\mu}\pr_jX^{\nu}\nonumber\\
{\rm and}\quad B_{ij}&=&{\hat B}_{\mu\nu}\
\pr_iX^{\mu}\pr_jX^{\nu}\ , \label{induce}
\eea
The gauge choice allows an independent treatment of the time component ($X^0$)
from the remaining spatial ones. On the boundary, it satisfies the 
Neumann condition $\pr_nX^0=0$, where `$n$' denotes the normal coordinate.
The B-field is a ($p\times p$)-matrix of rank $r=p/2$ and is magnetic
due to the gauge choice. In fact, the B-field can not be gauged
away in presence of a $D$ p-brane and the gauge invariance can be
maintained in combination with the $U(1)$ gauge field in the theory.
Since a $D$ p-brane is defined with Dirichlet boundary conditions,
$\d X^{\a}=0$, along the transverse directions ($\a=(p+1),\dots 25$),
the effective dynamics (\ref{openaction}) turns out to be the one on
its world-volume. Then the induce world-volume fields (\ref{induce})
can be identified with the longitudinal components ($h_{ij}={\hat G}_{ij}$
and $B_{ij}={\hat B}_{ij}$) of the background fields. Also, the gauge
field on the world-volume can be identified with the longitudinal
components of the $U(1)$ field.

\sp
\hs
The general nature of the world-volume fields is very difficult to treat
and one needs to adopt an approximation. We consider geodesic expansions 
for the fields ($h_{ij}$, $B_{ij}$ and $A_{i}$) around their zero modes and
they can be given by
\bea
{A}_i(X)&=& - \ {1\over2}{F}_{ij}\
X^j \ + \ \co ( {\pr F}) \ ,\nonumber \\
{\hat G}_{ij}(X)&=& \d_{ij} \ +\ \pr_{\l}{\hat G}_{ij}\ X^{\l} \ +\
\co({\pr^2{\hat G}}) \nonumber\\
{\rm and}\qquad {\hat B}_{ij}(X)&=& \ce_{ij}\ +\
\pr_{\l}{\hat B}_{ij} \ X^{\l}\ +\ \co({\pr^2{\hat B}}) \ .
\label{geogauge}
\eea
For slowly varying world-volume fields,
the derivative corrections
(\ref{geogauge}) can be ignored and the effective action (\ref{openaction})
for the world-volume dynamics (with a diagonal metric) becomes
\be
S \simeq\ {1\over{4\pi\a'}} \left [ \ \int_{\S} 
h_{ij}{\pr_{\bar a}} X^{i}
{\pr^{\bar a}} X^{j}
\ + \ (B + 2\pi\a'\  F)_{ij} {\int}_{\pr {\S}}  
X^{i}\pr_{\tau} X^{j} \ \right ] \ .\label{action}
\ee
The field strength ($B+2\pi\a' F$) is a gauge invariant
combination and thus the gauge field can be absorbed in the B-field
($\bar B=2\pi\a' B$). 
Since a $D$ p-brane carries charges ($Q_p$) coupled to the Ramond-Ramond (RR)
forms $C_{p+1}$, the Wess-Zumino (WZ) action for a generalized
$D$ p-brane can be symbolically given by
\be
S_{WZ}\ = \ Q_p\ \int \ dt d^p\s \ e^{\bar B}\wedge C \ . \label{wz}
\ee
Now the Neumann boundary conditions for a $D$ p-brane are derived from the 
effective action (\ref{action}). They are modified due to the
B-field and can be expressed as
\be
h_{ij}\ \pr_nX^j \ +\ {\bar B}_{ij}\ \pr_tX^j\ = 0 \ ,
\label{bcond}
\ee
where `$t$' denotes the tangential coordinate.

\subsection{Non-commutativity on the world-volume}

In this section, we present the canonical analysis to obtain the
non-commutativity among the $D$ p-brane collective coordinates.
Consider the
topology of the world-sheet as $\ci_{[0,\pi ]}\times \car$
(an infinite strip of width $0\leq\s\leq\pi$).
Then the space-like parameter, $\s$, runs along the length of the string and
$\tau$ can be considered as time-like.
The momenta conjugate to the coordinates can be given by
\be
P_i \ = h_{ij}\ \pr_t X^j + \ {\bar B}_{ij}\ \pr_n X^j 
\label{momenta}
\ee
In presence of B-field, the momenta (\ref{momenta}) 
receive correction and is known to be responsible for the
non-commutative description on the $D$ p-brane world-volume. 
The commutation relation between the conjugate
coordinates at the boundary can be derived (\ref{bcond}) from
that in the bulk among the canonical variables with an invertible
B-field. In general, the commutators on the boundary 
($\s=0$ and $\pi$) are subtle to define. In a time ($\tau $)
ordered space, the commutators can be expressed as
\bea
&& \Big [ X_L^i(\tau )\; ,\; X_L^j(\tau') \Big ] \ 
=\ i \pi \left (\ 2\Theta_S^{ij} \ -\ \Theta_A^{ij}\ \right )
\ce(\tau -\tau') \nonumber \\
{\rm and}\quad &&
\Big [ X_R^i(\tau )\; ,\; X_R^j(\tau')\Big ] \ =\ i \pi \left (\ 
2 \Theta_S^{ij}\ +\ \Theta_A^{ij}\right )\ce(\tau -\tau')\ ,
\label{nc}
\eea
where $X_L^i=X^i(\tau, 0)$ and $X_R^i=X^i(\tau, \pi )$ are the open string
operators defined, respectively, in the left and right ends of the
open string $X^i(\tau, \s)$ on the $D$ p-brane world-volume. 
Under the reversal of the prametrization, $\s \leftrightarrow \pi - \s$,
the  corresponding non-zero modes in the 
Fourier expansion for the operators ($X_L\leftrightarrow X_R$)
can be seen to remain invariant.
The right hand sides (\ref{nc}) for the commutators are $c$-numbers
and contain a symmetric ($i\leftrightarrow j$) $\Theta_S^{ij}$ 
and an antisymmetric $\T^{ij}$ matrix parameters.
Explicitly, they can be expressed as
\be
\Theta_S^{ij}= \left ({h\over{h+{\bar B}}}\cdot{h\over{h -
{\bar B}}}\right )^{ij} \quad 
{\rm and}\qquad\Theta_A^{ij} = \left ({1\over{h+{\bar B}}}{\bar B}{1\over{h - 
{\bar B}}}\right )^{ij} \ .\label{parameter}
\ee

\sp
\hs
The left and the right ends of the open string are separated by a
geodesic of string length and are out side each other light cone.
This in turn implies that no physical effect can be propagated between
the left and right space-like modules ($\ca_L$ and $\ca_R$).
The commutators between them become
\be
\left [ X_L^i(\tau )\; , \; X_R^j(\tau )\right ] \ =\ 0 \ .
\ee

\sp
\hs
The expression for the commutators (\ref{nc}) involve a discrete
value function $\ce(y)$ which takes $+1$ or $-1$, respectively for positive
and negative $y$. In a general description, 
$\ce(y)$ ensures the ordering in the phase space. 
A close observation on the non-commutativity (\ref{nc}) indicates that
the diagonal elements, $\Theta_S^{ij}$, are ordered in time, $\tau$,
where as the off-diagonal elements $\Theta_A^{ij}$ can be
ordered with respect to 
the space-like coordinates $X^i(\tau )$. To avoid the naive ambiguity 
(if any), while defining the commutators (\ref{nc}) at equal time
($\tau =\tau'$), one can alternately define the operators, $X^i(\tau )$, 
with a time ordering, $\tau>\tau'$, such that $|\tau -\tau'|$ is 
infinitesimally small $\eps$. As a result, the non-commutativity
in the space-like directions, $i\neq j$, can be interpreted intuitively as
a time ordered space. Then the 
commutators (\ref{nc}) in the left and right modules{\footnote{Henceforth,
we drop the index $L$ and $R$, since the left module is a commutant of
the right module and vice-versa. Physics is identical in both the modules
except for a signature (\ref{nc}).}} can be expressed, 
\be
\Big [ X^i(\tau )\; ,\; X^j(\tau) \Big ] \
=\ T\left [ X^i(\tau)X^j({\tau'_-})\ -\ X^i(\tau)X^j({\tau'_+})
\right ] \ .
\label{torder}
\ee

\sp
\hs Since the world-volume dynamics involves open string modules,
it describes a phase space for a $D$ p-brane.
The algebra $\ca$ of functions (in left and right modules) are 
generated by the operators $\cu^i=\exp (iX^i)$
and satisfy the $C^{\star}$ algebra{\footnote{$e.g.\ U^i(X)\star
\ U^j(X)= \exp (i\Theta_A^{ij}
\pr_{y^i}\pr_{z^j})\ U^i(X+y) U^j(X+z)|_{y=0=z}$. The $\star$-product
is associative and leads to Moyal bracket description.}} with
multiplication of functions on the non-commutative space.
It can be given by
\be
\cu^i\cdot \cu^j \ \ra \;
\cu^i\star\ \cu^j\ = \ e^{-\pi i\ \left (2\Theta_S^{ij} \mp \Theta_A^{ij}
\right )} \cu^i\cdot \cu^j
\ .\label{star}
\ee

\subsection{Effective non-commutativity: A Renormalized D-brane }

In this section, we discuss the effective non-commutativity 
as seen in a physical process such as in the scattering phenomena. 
For simplicity, consider the
tachyon emission vertices $V_n(k_n,\tau_n) = \exp \big [ik_n\cdot 
X(\tau_n\big ]$ for ($n=1,2\dots l$) inserted on the boundary. 
The scattering amplitude can be obtained by taking the 
expectation value for the time ordered ($\tau_1>\tau_2>\dots >\tau_l$) 
vertices:
\be
{\Big \langle} \prod_{n=1}^l V_n {\Big \rangle}_{\Theta_S,\Theta_A}= 
\exp \left (i\pi \sum_{n>n'} \left [ 2\Theta_S^{ij}k_{in}k_{jn'}
\pm \Theta_A^{ij}k_{in}k_{jn'}\right ]\ce (\tau_n -\tau_{n'}) \right )
{\Big \langle} \prod_{n=1}^l V_n {\Big \rangle}_{\Theta_A=0} \ .\label{vertex}
\ee
Using momentum conservation, $\sum_nk_n = 0$, it can be checked that the
symmetric part $\Theta_S^{ij}$ in the 
phase factor (\ref{vertex}) does not contribute substantially.
Thus the effective non-commutative parameter is 
due to the antisymmetric matrix elements $\Theta_A^{ij}$ which in turn 
implies the presence of B-field. Then
the effective phase responsible for the world-volume non-commutativity 
reduces to
\be
\exp \left (\pm 2i\pi \sum_{n>n'}\Theta_A^{ij}k_{in}k_{jn'}\ce (\tau_n -
\tau_{n'} )\right ) \ . \label{effective}
\ee
This expression implies that the non-commutativity on the world-volume
is solely due to the broken phase of cyclic symmetry in the time ordered
space. This in turn implies that the antisymmetric parameter, $\Theta_A^{ij}$,
plays a vital role for the non-commutative description on the world-volume.

\sp
\hs
The illustration (\ref{effective}) shows that the world-volume fields
should be re-defined to absorb the symmetric matrix elements,
$\Theta_S^{ij}$. In fact, the diagonal matrix elements are
associated with a short distance divergence
\cite{kk,kar98,kar99}, which can be seen while defining the correlators in 
sec. 3.1. The divergence can be regulated by introducing a cut-off. This
in turn implies that the world-volume should be properly re-normalized
to absorb the symmetric matrix elements completely. If ${\bar X}^i$
denote the renormalized coordinates
\be
X^i(\tau )= \ {\bar X}^i(\tau )\ + \ \d {\bar X}^i(\tau )\ ,
\label{renorm}
\ee
then $\d {\bar X}^i$ contain the divergence and can be regulated by
introducing a cut-off.
The coefficient of the divergence piece (\ref{renorm}) is 
symmetric in ($i\leftrightarrow j$) and can be 
intuitively obtained from ref.\cite{kar99}
\be
\d {\bar X}\ \approx \ \left ( \Theta_S \ +\ \Theta_{S'}
\right )\  K \cdot [{\rm divergence}] \ , \label{ren}        
\ee
where the extrinsic curvature $K$ can be obtained
by considering the Lie derivative on the induced fields ($h,{\bar B}$).
The additional symmetric elements, $\Theta_{S'}^{ij}$, can be expressed as
\be
\Theta_{S'}^{ij}\ = \ \left ( {{\bar B}\over{h+{\bar B}}}
\cdot{{\bar B}\over{h-{\bar B}}}\right )^{ij} \ .
\ee
As a result (\ref{renorm}),
the world-volume fields (\ref{induce}) become renormalized 
$i.e.\ h\ra {\bar h}$ and $B\ra {\bar B}$.{\footnote{The detail
of the computation for the re-normalization can be obtained from 
ref.{\cite{kar99}) by a suitable conformal transformation.}}
Then the commutators (\ref{nc})
in terms of the renormalized coordinates in the time ordered space become
\be
\Big [ {\bar X}^i(\tau )\; ,\; {\bar X}^j(\tau') \Big ] \
=\ \mp i \pi {\Theta}_A^{ij}\  \ce(\tau -\tau') \ . \label{rnc}
\ee
Since the short distance divergence is renormalized, one can also,
alternately, re-write the
commutators (\ref{rnc}) at equal time ($\tau=\tau'$) where ordering
is implicit among the space-like coordinates ${\bar X}^i(\tau)$. 
For the renormalized coordinates, the commutators
takes the form:
\be
\Big [ {\bar X}^i(\tau )\; ,\; {\bar X}^j(\tau) \Big ] \
=\ \mp i \pi {\Theta}_A^{ij}
\ .\label{rnc2}
\ee

\sp
\hs
On the other hand, the boundary interaction (\ref{action}) naively seems 
to be invariant under an infinitesimal $U(1)$ gauge transformation
$\delta A_i=\pr_i \eps$. However, the gauge field $A_i(X)$ itself possesses
a dependence on the non-commuting coordinates. Thus a proper treatment of
the gauge field is necessary to account for the string coordinates. In fact,
such an approximation (geodesic expansion around its zero modes) has
already been taken into account (\ref{action}) for slowly varying B-field.
As a result the $U(1)$ gauge sector is constrained by the
collective coordinates ${\bar A}_i\simeq {-{1\over2}{\bar B}_{ij}{\bar X}^j}$.
This in turn implies that the gauge transformation can be viewed
as a re-parameterization invariance
\be
\d{\bar A}_i\ = \ - {1\over2}{\bar B}_{ij}\ \d {\bar X}^j \ + \co(\pr B).
\ee
For the renormalized $D$ p-brane world-volume, the
equal-time commutator in the Abelian ($\l$ being its generator) gauge
sector becomes
\be
\Big [ {\bar A}_i \ \; , \; {\bar A}_j \Big ] \ = \ \pm {{i\pi\l^2}\over4}
\ \Theta_A^{kk'}{\bar B}_{ik} {\bar B}_{k'j}\ + \co(\Theta_A^2)\ .
\label{gaugecom}
\ee
\section{Geometry and zero modes}

\subsection{World-volume correlators}

Now let us study correlation function defined with the operators, 
$X^i(\tau )$ in the left and right modules 
on the $D$ p-brane world-volume. They define the
Neumann propagator, $G^{ij}(\tau ,\tau ') = \langle X^i
(\tau )X^j(\tau ') \rangle$, on the world-volume which is a
($p\times p$)-matrix valued.
The propagators satisfy the boundary condition (\ref{bcond})
\be
h_{ij}\ \pr_nG^{jk}\ + \ i\ {\bar B}_{ij}\ \pr_tG^{jk}
\ =\ 0 \ .\label{bprop}
\ee
The explicit expression for the 
propagator $G^{ij}(\tau,\tau' )$ 
on the world-volume can also be obtained by a conformal transformation
from our earlier work \cite{kar99}. The boundary value propagator defines
the world-volume correlators and can be given by
\be
G^{ij}(\tau ,\tau ')=\ 2 \ \Theta_S^{ij} \ \ln |\tau -\tau'|
\ \mp \ {{i\pi}\over2}\ \Theta_A^{ij}\ \ce (\tau -\tau') \ .
\label{prop}
\ee
As discussed
in section 2.2, the limit $\tau\ra\tau'$ can 
be considered with a time ordering $\tau>\tau'$ and 
when $|\tau -\tau'|$ is kept infinitesimally small. Then the propagator
in the limit, $\eps\ra 0^{\pm}$, becomes
\be
G^{ij}(\tau ,\tau )=\ 2 \ \Theta_S^{ij} \ \ln \eps 
\ \mp \ {{i\pi}\over2}\ \Theta_A^{ij}\ .
\label{prop2}
\ee
It shows that in presence of B-field,{\footnote{If $B=0$, 
the propagators (\ref{prop})
are well defined and can be seen to contain a logarithmic divergence $\ln\eps$.
As a standard practice, the loop divergences
contained in $\langle X^i(\tau )X^i(\tau)\rangle $ can be regulated by
Pauli-Villars regularization scheme. In this approach, one renormalizes the
world-volume fields (\ref{renorm}) at the expense of the vertex operators.
Then every element in the
propagator precisely corresponds to that of a $D$-particle.}}
a finite contribution is associated with the 
off-diagonal matrix elements, $\Theta_A^{ij}$, which is antisymmetric.
For $\tau=\tau'$ in eq.(\ref{prop}), 
the the symmetric elements, $\Theta_S^{ij}$,
are associated with a logarithmic divergence and 
the antisymmetric elements
are accompanied with a discrete value function 
$\ce(\eps )$ which is not well defined there. 
Then, a priori, the Pauli-Villars regularization does not seem to be 
appropriate in presence of B-field. However in the limit, 
$\tau\rightarrow\tau'$,
the propagator (\ref{prop}) is well behaved which essentially defines
the point-splitting regularization. As a result,
different operators ($i\neq j$) are not allowed at the
same time ($\tau\neq \tau'$). It is necessarily due the antisymmetric
property of B-field which defines the point-splitting scheme.

\sp
Once we obtain the necessary re-normalization for the
$D$ p-brane world-volume (\ref{ren}), the symmetric elements can be
gauged away in a physical process (\ref{effective}). Then the re-normalized
propagator can be seen to be constant (anti-symmetric) matrix and 
re-written as
\be
{\bar G}^{ij}(\tau ,\tau ')= \ \mp \ {{i\pi}\over2}
\ \Theta_A^{ij}\ \ce (\tau -\tau' )\ , \label{renprop}
\ee
In principle,
the renormalized $D$ p-brane dynamics \cite{kar99} receive 
$\a'$-corrections (due to the curvatures) associated with the
antisymmetric matrix elements $\Theta_A^{ij}$. The higher order terms 
associated with $\a'$ can be argued for the corrections to the commutative
geometry. This in turn implies that the $\a'$ corrections (accompanied with
the antisymmetric property) correspond to the deformation of the Yang-Mills
theory on the world-volume and
lead to an associative algebra in the left  
and right modules. The simple analysis explains 
two different geometrical aspects for a generalized $D$ p-brane which
is in agreement with the recent analysis \cite{sw}. 

\subsection{SO(p) symmetry: Homomorphic image space} 

\hs
In general, a rotation among the spatial coordinates changes the
topology leading to equivalent geometries.
For instance, under a
rotation, a (diagonal) metric and an anti-symmetric two-form field can be 
transformed to anisotropic geometry in presence of a new metric and vanishing
two-form field. As a result, the non-commutative geometry can be viewed, 
alternately, with a non-local description on the (rotated) $D$ p-brane
world-volume. In
other words, a re-normalization, in the context, leads to a world-volume
re-definition and can be seen as a rotation among the (spatial) collective 
coordinates defining the $D$ p-brane.

\sp
\hs
In fact the constant matrix coefficients in the regulated
propagator (\ref{renprop}) correspond to a rotation in the
Cartan sub-algebra. Then the world-volume can be described 
in terms of $p/2$ independent blocks. 
The metric
and the B-field can be expressed as
\be
h_{ij}\ = \ \oplus_{r=1}^{p/2}\pmatrix {
{h_r} & {0}\cr {0} & {h_r} \cr } \qquad {\rm and}\quad
B_{ij}\ =\ \oplus_{r=1}^{p/2}\pmatrix {
{0} & {B_r}\cr {-B_r} & {0} \cr } \ . \label{2dim}
\ee

\sp
\hs
Now consider the zero modes on the world-volume, $x^i=X^i-\xi^i$, which
can be seen to play an important role in presence of B-field.
These (constant) modes are characterized by a group of translations
on the world-volume and is responsible for the inhomogeneous space.
On the other hand, the non-zero modes are described by a group of
rotation, $SO(p)$, for a $D$ p-brane. Under an inhomogeneous transformation
\be
X^i\ \rightarrow R^{ij}\xi_j\ + \ x^i \ ,\label{trans}
\ee
where $R^{ij}$ denotes the rotation matrix \cite{kar99} and
can be written in the ($2\times 2$)-block
diagonal form:
\be
R^{ij}= \ \oplus_{r=1}^{p/2}
\pmatrix { {{\Theta_S^r} + \Theta_{S'}^r} & {2 i\ \Theta_A^r} \cr
{-2i\ \Theta_A^r} & {{\Theta_S}^r + \Theta_{S'}^r} \cr } \ . \label{rotation}
\ee
It is little tricky to define the corresponding rotation angles, $\beta^r$, 
as the metric is not flat. Finally, the angle for each ($2\times 2$)-block
can be written as 
\bea
\beta^r&=&\sin^{-1}\left (2i\ \Theta_A^r\right ) \ ,\nonumber\\
&=& \cos^{-1}\left( \Theta_S^r \ +\ \Theta_{S'}^r\right ) \ .
\eea
Now the world-volume geometry can be analyzed from that of
a real plane.{\footnote{Henceforth, we omit the index $r$ unless it is 
required.}} The new coordinates, $X_{\pm}(\tau )=x_{\pm} + \xi_{\pm}$,
describe a two dimensional phase space
and can be given by
\bea
\xi_{+}&=& \xi^1 \ \sin\beta \ + \ \xi^2\ \cos\beta \nonumber \\
\xi_{-}&=& \xi^1 \ \cos\beta \ -\ \xi^2\ \sin\beta \ .
\label{twist}
\eea
Since the rotation angle, $\beta$, is not fixed, it gives rise to
a non-local description on the $D$ p-brane world-volume. 

\sp
\hs
In the new frame, 
the boundary condition (\ref{bcond}) turns out to be
\bea
&& {D^{\pm}}{\xi_{\pm}}(\tau ) \ =0\ ,\nonumber\\
{\rm where}\quad &&
{D^{\pm}}= (h\ \pr_n \pm\ i B\ \pr_t ) \ .
\eea
The resultant directions ($\pm$)
intertwine the Neumann and Dirichlet conditions
\cite{kar99}. The nature of the differential, $D^{\pm}$, is
determined by the B-field.

\sp
\hs
For a re-normalized $D$ p-brane world-volume,
the  commutator (\ref{rnc2}) in the new frame
can be given by
\be
\Big [{\bar X}_{-}^r(\tau )\;  ,\; {\bar X}_{+}^r(\tau)\Big ] \ =\ \mp i\pi\
\Theta_A^{r} \ . \label{twistnc}
\ee
Since the presence of zero modes make the world-volume
an inhomogeneous space, the translation group does not commute
with the rotation $SO(p)$. Then the commutator (\ref{twistnc}) can be
expressed in terms of the zero and non-zero modes of the un-twisted
world-volume :
\be
\Big [{\bar X}_{-}\;  ,\; {\bar X}_{+}\Big ] \ =\ 
\Big [{\bar x}^{1}\;  ,\; {\bar x}^{2} \Big ] +
\Big [{\bar x}^{1}\;  ,\; {\bar\xi}^{2}\Big ] +
\Big [{\bar\xi}^{1}\;  ,\; {\bar x}^{2}\Big ] +
\Big [{\bar\xi}^{1}\;  ,\; {\bar\xi}^{2}\Big ] \ .
\label{zerocom}
\ee
All four of the commutators (\ref{zerocom}) contribute (independently)
towards the non-commutativity on the world-volume. Under a
homomorphic transformation, the zero modes can be modded out to obtain
a homogeneous description on a $D$ p-brane. As a result, the homomorphic
image space is commutative among the zero and non-zero modes though the
world-volume retains the ordering (independently) 
among the zero and non-zero mode sectors. Then the commutator (\ref{twistnc})
can be given by
\be
\Big [{\bar x}_{-}\;  ,\; {\bar x}_{+} \Big ]\  +\
\Big [{\bar\xi}_{-}\;  ,\; {\bar\xi}_{+}\Big ]\ =\ \mp i\pi\ 
\Theta_A\ .
\label{ncom3}
\ee
The analysis shows that
the non-commutative symmetry can be mapped onto a commutative one under a
homomorphic transformation. Thus the
$U(1)$ sector of the world-volume describes the homomorphic image
space for the non-commutative space which is described by the 
$SO(p)$ rotation group.

\sp
\hs
In this frame, the constant re-normalized matrix propagator (\ref{renprop}) 
turns out to be diagonal and satisfies
\be
D^{\pm}{\bar G}_{D}(\tau,\tau' )\ = 0 \ .
\ee
Explicit expression for the propagator can be given by
\be
{\bar G}_D(\tau, \tau') 
\ = \pmatrix { {{1\over2}\pi \Theta_A } & {0} \cr {0} & 
{- {1\over2}\pi \Theta_A } \cr }\ \ce (\tau -\tau ') \ . \label{rotprop}
\ee
The diagonal form of the re-normalized propagator  
confirms (following the discussions in sec.2.3) the
commutative aspect of a rotated $D$ p-brane world-volume.
In the limit $\tau\rightarrow\tau'$, the propagator (\ref{rotprop})
contributes a finite constant (with antisymmetric property).
This in turn,  can be seen to introduce a constant shift on the world-volume
fields leading to 
quantum description on a $D$ p-brane world-volume. 

\subsection{Strong magnetic field: Zero modes non-commutativity}

Consider a limiting case for the B-field, 
$i.e.$ strong B-field
or large B limit ($B>> 1$). 
As discussed in the current literatures \cite{sw,ishibashi,okuyama},
the limit can be seen to shed light on the non-commutative aspect
of a $D$ p-brane as the analysis becomes simpler there.
In the limit, the leading term from the WZ-action
(\ref{wz}) becomes that for a large density ($\omega$)
of $D$-particles. The leading WZ-term in the limit becomes
\bea
&&S_{WZ}^{D_0}\simeq \ Q_0 \ \omega \ \int \ dt \ C_1 \ ,\nonumber\\
{\rm where}\quad &&\omega \ = {1\over{2\pi}} \int\ d^p\s \ B^p \ .
\label{largeB}\eea
This is the idea of branes within higher branes
\cite{douglas}. It can be seen as 
the building block for the M-theory description of matrix
model \cite{bfss} as well as type IIB matrix theory \cite{ikkt}.

\sp
\hs
For large B, (the induced metric $|h|<<1$), the
leading order description on the world-volume turns out to be topological.
In the limit, the world-volume is in agreement with the holographic principle.
As a result, the 
closed string dynamics in the bulk can be reproduced in terms of the
open string dynamics at its
boundary. This suggests that the world-volume non-commutative geometry
can be considered
equivalent to the closed string (commutative) 
geometry in the bulk.{\footnote{In fact, the closed
string channel is T-dual to the open string channel. In 
closed string sector, the left and right modules overlaps due to the 
periodicity in its boundary value and the non-commutative description 
(\ref{rnc}) cancels against each other.}}
This in turn implies that
the non-commutative Yang-Mills dynamics on the $D$ p-brane world-volume
should be equivalent to the commutative dynamics in the closed string
sector.

\sp
\hs
In the limit,
the boundary conditions (\ref{bcond}) becomes
\be
\pr_tX^i\ = \ 0 \ . \label{strongB}
\ee
It shows that the
string modes, $X^i$, are identified with the zero modes, $x^i$, in the 
large B limit. The expression (\ref{strongB})
corresponds to the Dirichlet condition for $D$-particles on a 
$D$ p-brane world-volume. This in turn implies that
the $D$ p-brane world-volume 
becomes spatially non-localized and represents a high density of
$D$-particles. In addition, the world-volume 
non-commutativity (\ref{nc}) can be seen only
among the zero modes, $x^i$, independently in the left and right modules. 

\sp
\hs
Here, the non-commutative parameter $\Theta_A^{ij}= - ({1/B})^{ij}$
and  the equal-time commutator (\ref{rnc2})
reduces to that 
among the zero modes of the collective coordinates ($X^i,{\bar A}_i$).
It takes the form:
\bea
&&\Big [{\bar x}^i \; , \; {\bar x}^j \Big ]=\ \pm i\pi \ 
\left ({1/B}\right )^{ij} \ ,\nonumber\\
&&\Big [{\bar x}^i \; , \; {\bar A}^j \Big ]=\ \mp 
{{i\pi}\over2}\ h^{ij}\ .
\eea
It shows that the collective coordinates
in the gauge sector becomes
the canonical momenta to the conjugate coordinates or vice-versa
in presence of a B-field. In general,
non-commutative description (\ref{nc}) holds good for any value
of B-field though in the limit, a clear picture of $D$-particles
described by the zero modes emerge out leaving the
feebly coupled $D$ p-brane. 

\sp
\hs
In the new frame, 
discussed section 3.2, the picture can be easily identified
with the motion of a large number (\ref{largeB}) of $D$-particles in an uniform 
and strong magnetic field (B-field). Since the $D$-particles are described
only by the zero modes, 
the wave function representing its ground state,
$\psi = \exp \left (
i\ {k}_{-}{\bar x}_{-}\right )$ for the momentum 
$k_{-}={\pi^{-1}}{\bar B}x_{+}$, is associated with
an additional phase due to the non-commutative description,
$[{\bar x}_{-},{\bar x}_{+}]=\pm i\pi B^{-1}$. 
It can be given by
\be
\psi \ |_{\rm (strong\ B)}=  
e^{\mp {i\over{\pi}}{\bar B}{\bar x}_{-}} \cdot \psi\  |_{(B=0)}
\ . \label{degenerate}
\ee
Since the world-volume is a compact space,  the single-value condition 
on the wave-function can be imposed by introducing a
periodicity in one of the spatial coordinate (${\bar x}_{-}$). Then
its conjugate momentum becomes (Dirac) quantized:
${k}_{-} = 2\pi n$ for an integer $n$ and the phase factor 
(\ref{degenerate}) reduces to: $\exp (2\pi i n\ {\bar x}_{-})$.
Since the Landau levels are independent of the momentum, ${k}_{-}$,
the $D$-particle wave function degenerates in presence of a strong magnetic
field leading to a non-local description. Though there are some
subtleties due to the boundary value of the coordinate field,
the illustration shows that
the non-commutative description leads to non-localization of $D$-particles
on the $D$ p-brane world-volume.
Then the non-commutative Yang-Mills symmetry on the world-volume may be 
better understood from the study of a simple physical model for
the $D$-particles.

\section{Discussions}

To summarize, we have shown that a re-normalization of the  
generalized $D$ p-brane is essential 
to account for a precise non-commutative geometry on its world-volume.
We learned that the B-field is essential, due to its antisymmetric
property, for the non-commutative aspect of a $D$ p-brane. Due to the
subtleties in defining the non-commutativity, we have considered a time
ordering among the operators as an alternative to the non-commutative 
description. The world-volume
geometry is analyzed with respect to the $SO(p)$ rotational symmetry group. The
zero modes were seen to introduce the non-commutative feature between the
group of translations and the rotation. A homomorphic map from the 
non-commutative space was argued for the Abelian gauge symmetry. Large
B limit was analyzed and the zero modes non-commutativity on the
$D$ p-brane world-volume were discussed. 
In the large B limit, the correspondence
between the non-commutative and commutative symmetries can be obtained by
implementing the Holographic idea. The existence of such a holographic
connection between different geometries may enhance our understanding of
quantum gravity. 

\sp
\hs
In the actual computations, though the
world-volume fields are slowly varying, the non-commutative parameter
matrix $\Theta_A^{ij}$ is assumed to be a constant for simplicity. In
principle, $\Theta_A^{ij}$ is not fixed. It would imply that the deformation
parameter should be expanded (a convergent series)
around its zero mode. However such a generalization is out of reach
in the present context. 

\vfil\eject

\noindent {\large\bf Acknowledgments}

\sp

\hs
I would like to thank M. Cederwall, G. Ferretti and B.E.W. Nilsson
for various useful discussions. The work is supported by the
Swedish Natural Science Research Council.
\sp
\sp
\def\anp{Ann. of Phys.}
\def\prl{Phys. Rev. Lett.}
\def\prd#1{{Phys. Rev.} {\bf D#1}}
\def\plb#1{{Phys. Lett.} {\bf B#1}}
\def\npb#1{{Nucl. Phys.} {\bf B#1}}
\def\mpl#1{{Mod. Phys. Lett} {\bf A#1}}
\def\ijmpa#1{{Int. J. Mod. Phys.} {\bf A#1}}
\def\rmp#1{{Rev. Mod. Phys.} {\bf 68#1}}



\begin{thebibliography}{99}

\bibitem{polchinski}J. Polchinski, {\prl } {\bf 75} (1995) 4724
({\tt hep-th/9510017}).

\bibitem{dkps}M. Douglas, D. Kabat, P. Pouliot and S. Shenker,
\npb{485} (1997) 85 \\
({\tt hep-th/9608024}).

\bibitem{bfss}T. Banks, W. Fischler, S.H. Shenker and L. Susskind
\prd{55} (1997) 5112 \\
({\tt hep-th/9610043}).

\bibitem{ikkt}N. Ishibashi, H. Kawai, Y. Kitazawa and A. Tsuchiya
\npb{498} (1997) 467\\
({\tt hep-th/9612115}).

\bibitem{connes}A. Connes, M.R. Douglas and A. Schwarz, J.H.E.P.
{\bf 2} (1998) 003 ({\tt hep-th/9711162}).

\bibitem{hull}M.R. Douglas and C. Hull, J.H.E.P. {\bf 9802} (1998)
008 ({\tt hep-th/9711165}).

\bibitem{sw}N. Sieberg and E. Witten, J.H.E.P. {\bf 9908} (1999)
({\tt hep-th/9908142}).

\bibitem{witten86}E. Witten, \npb{268} (1986) 253.

\bibitem{schomerus}V. Schomerus, J.H.E.P. {\bf 9906} (1999) 030
({\tt hep-th/9903205}).

\bibitem{berkooz}M. Berkooz, \plb{430} (1998) 237 ({\tt hep-th/9802069}).

\bibitem{krogh}Y.-K. E. Cheung and M. Krogh, \npb{528} (1998) 185
({\tt hep-th/9803031}).

\bibitem{kawano}T. Kawano and K. Okuyama, \plb{433} (1998) 29
({\tt hep-th/9803044}).

\bibitem{ardalan}F. Ardalan, H. Arfaei and M. M. Sheikh-Jabbari,
{\tt hep-th/9803067};  J.H.E.P. {\bf 9902} (1999) 016 
({\tt hep-th/9810072}); {\tt hep-th/9906161}.

\bibitem{ho}P.-M. Ho, \plb{434} (1998) 41 ({\tt hep-th/9803166}).


\bibitem{chu}C.-S. Chu and P.-M. Ho, 
\npb{550} (1991) 151 ({\tt hep-th/9812219});
({\tt hep-th/9906192}).

\bibitem{kato}M. Kato and T. Kuroki, J.H.E.P. {\bf 9903} (1999)
({\tt hep-th/9902004}).

\bibitem{kar99}S. Kar, {\tt hep-th/9907117}.

\bibitem{susskind}D. Bigatti and L. Susskind, {\tt hep-th/9908056}.

\bibitem{li-Wu}M. Li and Y-S. Wu, {\tt hep-th/9909085}.

\bibitem{ishibashi}N. Ishibashi, {\tt hep-th/9909176}.

\bibitem{okuyama}K. Okuyama, {\tt hep-th/9910138}.
 
\bibitem{tseytlin}E.S. Fradkin and A.A. Tseytlin, \plb{163}
(1985) 123.

\bibitem{callan}A. Abouelsaood, C.G. Callan, C.R. Nappi and
S. A. Yost, \npb{280} [FS18] (1987) 599;
C. G. Callan, C. Lovelace, C. R. Nappi and S. A. Yost,
\npb{288} (1987) 525.

\bibitem{kk}S. Kar and Y. Kazama, \ijmpa{14} (1999) 1531
({\tt hep-th/9807239}).

\bibitem{kar98}S. Kar, \npb{554} (1999) 163 ({\tt hep-th/9812230}).

\bibitem{douglas}M. R. Douglas, {\tt hep-th/9512077}.

\end{thebibliography}
\end{document}